\documentclass[twocolumn]{aastex6}

\usepackage{graphicx}
\usepackage{amssymb}
\usepackage{amsmath}

\usepackage{natbib}
\usepackage{txfonts}
\usepackage{multirow}
\usepackage{array}
\usepackage{sidecap}
\usepackage{hyperref}
\usepackage{epstopdf}
\usepackage{footnote}
\usepackage{tabularx}
\usepackage{booktabs}
\usepackage{longtable}
\usepackage{tabu}

\newcommand{\kms}{km~s$^{-1}$}

\newcommand{\Rsun}{$R_\odot$}

\newcommand{\Hinode}{\textit{Hinode}}

\newcommand{\sdo}{\textit{SDO}}

\shortauthors{Panesar et al.}

\begin{document}
 \title{\bf Magnetic Flux Cancelation as the Trigger of Solar Coronal-Hole Coronal Jets}

\author{Navdeep K. Panesar\altaffilmark{1}, Alphonse C. Sterling\altaffilmark{1}, Ronald L. Moore\altaffilmark{1,2}}
\affil{{$^1$}NASA Marshall Space Flight Center, Huntsville, AL 35812, USA}
\affil{{$^2$}Center for Space Plasma and Aeronomic Research (CSPAR), UAH, Huntsville, AL 35805, USA}
\email{navdeep.k.panesar@nasa.gov}


\begin{abstract}
	We investigate in detail the magnetic cause of  minifilament eruptions that drive coronal-hole jets.  We study 13 random on-disk coronal hole jet eruptions, using high resolution X-ray images from \Hinode/XRT, EUV images from \sdo/AIA, and magnetograms from \sdo/HMI. For all 13 events, we track the evolution of the jet-base region and find that a \textit{minifilament} of cool (transition-region-temperature)  plasma  is present prior to each jet eruption. HMI magnetograms show that the minifilaments reside along a magnetic neutral line between majority-polarity and minority-polarity magnetic flux patches.  These  patches converge and cancel with each other, with an average cancelation rate of  $\sim$ 0.6 $\times$ 10$^{18}$ Mx hr$^{-1}$ for all 13 jets. Persistent flux cancelation at the neutral line eventually destabilizes the minifilament field, which erupts outward and produces the jet spire. Thus, we find that all 13 coronal-hole-jet-driving minifilament eruptions are triggered by flux cancelation at the neutral line. These results are in agreement with our recent findings  \citep{panesar16b} for quiet-region jets, where flux cancelation at the underlying neutral line triggers
	the minifilament eruption that drives each jet. Thus from that study of quiet-Sun jets and this study of coronal hole jets, we conclude that flux cancelation is the main candidate for triggering quiet region and coronal hole jets. 

\end{abstract}

\section{INTRODUCTION}

Solar coronal jets are dynamic, evanescent and beam-like structures that often appear bright in coronal images \citep{raouafi16}. Regardless of the solar cycle jets are a pervasive solar phenomena, with some reaching heights of several hundred km in EUV and X-ray images \citep{shimojo96,savcheva07,shibata07}. White light jets, at least some of which are correlated with coronal jets, can reach several \Rsun\ \citep[e.g.][]{wangym98,ko05}. They occur in all types of solar environments: active regions \citep{shibata92,panesar16a,sterling16}, quiet regions \citep{innes16,panesar16b} and coronal holes \citep{nistico09,sterling15}. Jets are more prominent and often seen best in coronal-hole regions because the jet plasma is observed in emission against a dark coronal background in EUV and X-ray images \citep{cirtain07,sterling15}. 


\floattable
\begin{table*}
	\caption{Timing and location for the observed coronal-hole jets \label{tab:list}}
	\renewcommand{\arraystretch}{1.0}
	\begin{tabular}{c*{10}{c}}
		\noalign{\smallskip}\tableline\tableline \noalign{\smallskip}
		Event  &  Date &  Time\tablenotemark{a}& Location\tablenotemark{b}  & Jet Speed\tablenotemark{c}   & Jet Dur.\tablenotemark{d} & Jet-Base\tablenotemark{e}   & XRT\tablenotemark{f}  & $\Phi$ values\tablenotemark{g}  & \% of $\Phi$\tablenotemark{h} \\
		
		No.   &     & (UT) &Helio. Cord. & (\kms) & (minutes) & Width (km) & Coverage & 10$^{19}$ Mx & Reduction  \\
		
		\noalign{\smallskip}\hline \noalign{\smallskip}
		J1   & 2012 Jul 02 & 02:11 & N21, W07 & 65$\pm$1.5 & 10$\pm$2 & 22500$\pm$1500 & No   &  1.5 &  33$\pm$4.5   \\ 
		
		J2   & 2015 Aug 18 &   13:07 & N27, E02 & 40$\pm$20 & 12$\pm$3  &9700$\pm$1000 &   Yes   &  0.9 &  31$\pm$10 \\
		
		J3   & 2015 Dec 28 & 09:54 &N36, E19  & 110$\pm$40 & 7$\pm$1  & 13000$\pm$900&  No & 0.9  &  48 $\pm$9.0   \\   
		
		J4   & 2015 Dec 28 & 16:02 & N37, E03 & 35$\pm$7  & 11$\pm$1  & 12000$\pm$2500&  No &1.2  &  52$\pm$5.0   \\ 
		
		J5   & 2015 Dec 30  & 15:14  & N36, W23 & 70$\pm$30 & 7$\pm$1   & 6700$\pm$1000& No & 0.7   &  43 $\pm$10    \\  
		
		J6   & 2015 Dec 31  & 19:04 & N43, W34 & 27$\pm$4 &  6$\pm$1 & 6600$\pm$500 &  No & 0.6  & 41 $\pm$8.5   \\  
		
		J7   & 2016 Jan 01  & 11:45 & N08, E30 & 30$\pm$5 & 8$\pm$1  & 18500$\pm$3500 &  No  & 0.7 & 52 $\pm$6.5 \\   
		
		J8   & 2016 Jan 01  & 18:11 & N41, E39 & 204$\pm$70 & 4$\pm$1  & 1200$\pm$500 &  No  & 0.5 &  37 $\pm$10  \\   
		
		J9  & 2016 Apr 21  & 06:15 & S01, E12 & 240$\pm$70 & 8$\pm$1\tablenotemark{i}  & 10500$\pm$700 & Yes   &  --\tablenotemark{j}   & --  \\
		
		J10  & 2016 Sep 15  &   23:36  & S06, E00 & --\tablenotemark{k} & 6$\pm$1\tablenotemark{l}  & 19000$\pm$2000 &  Yes   &  --\tablenotemark{m}  & --  \\ 
		
		J11   & 2017 Jan 03   & 14:59 & N20, E02 & 105$\pm$30  & 6$\pm$1  & 18700$\pm$6000 & Yes & 1.6 & 33$\pm$5.5   \\ 
		
		J12   & 2017 Jan 04  & 09:26  & N24, W03 & 92$\pm$30  & 15$\pm$3  & 9500$\pm$1000 & Yes   & 2.0  &   73 $\pm$5.0  \\ 
		
		J13  & 2017 Jan 04  & 17:08  & N19, W10 & 103$\pm$30 & 4$\pm$1  & 7000$\pm$500 &  Yes   &  0.9   &  21$\pm$8.0  \\

		\noalign{\smallskip}\tableline\tableline \noalign{\smallskip}
		
	\end{tabular}
	
	\tablenotetext{a}{Approximate time of JBP brightening in AIA 94 \AA\ images.}
	\tablenotetext{b}{Approximate location of the jet-base region during the eruption onset.} 
	\tablenotetext{c}{Plane-of-sky speed along the jet spire. Speeds and uncertainties are measured from  AIA 171\AA\ time-distance maps.}
	\tablenotetext{d}{Duration of jet spire visibility in 211 \AA\ images.}
	
	\tablenotetext{e}{Measured within 30--60 minutes before the jet eruption.}
	\tablenotetext{f}{\Hinode/XRT jet data coverage. }
	\tablenotetext{g}{ Average flux ($\Phi$) values of the minority flux clumps 3-4 hours before jet eruption.}
	\tablenotetext{h}{Flux  change between 3-4 hours before jet eruption and 0-2 hours after eruption.}
	\tablenotetext{i}{Spire is clearly visible in XRT images but not visible in the AIA images. The given duration and speed is measured from the XRT images. }
	\tablenotetext{j, m}  {~~~ Magnetic field is too weak to measure but cancelation is clearly visible at the neutral line.}
	\tablenotetext{k}{Spire is too faint to measure the speed.}
	\tablenotetext{l}{ The given duration is based on both XRT and AIA  images.}
	
\end{table*}


Recent observations by \cite{sterling15} show that polar coronal-hole jets are driven by \textit{minifilament} eruptions, and a jet bright point (JBP) appears at the location from where the minifilament erupted. Later, \cite{panesar16b} studied 10 on-disk quiet-region jets and found that the same minifilament-eruption idea also holds for coronal jet generation in quiet-regions. They also investigated the triggering  mechanism of their quiet Sun coronal jets, and found strong evidence that progressive magnetic flux cancelation at a neutral line under the minifilament between  majority-flux and minority-flux clumps eventually destabilizes the field holding the minifilament material, resulting in an outward eruption of the minifilament. In each of the 10 events \cite{panesar16b} found clear evidence of ongoing flux cancelation at the neutral line. 

Very recently, \cite{panesar17} investigated the formation mechanism of quiet region pre-jet minifilaments that erupted to generate the jets that they analyzed in \cite{panesar16b}. In  \citep{panesar17}, they mainly investigated the longer-term magnetic field evolution that led to the formation of the pre-jet minifilaments. They found that oppositely-polarity flux patches converged and partially canceled, before and during the formation of the minifilaments over the neutral line between the converging flux patches. Continuous flux cancelation over several hours resulted in thickening and increased prominence of the  minifilaments at the neutral line. Eventually, flux cancelation between the oppositely-polarity flux patches triggered each minifilament eruption, leading to the jet. In summary, from our previous study of quiet-region jets \citep{panesar16b,panesar17}, we infer that in quiet regions 
magnetic flux cancelation is the main process for the build-up of the sheared and twisted field of the pre-jet minifilament, and is also the trigger of the minifilament's  eruption that makes the jet. 
In addition, \cite{sterling16,sterling17} found that, at least in most cases they examined, flux cancelation resulted in active region jets.

Because \cite{sterling15} studied polar coronal-hole jets, their events were too close to the polar limb for reliable magnetic field investigations. 
Thus, from the study of polar coronal-hole jets by \cite{sterling15}, the question of whether coronal-hole jets follow the flux cancelation idea (that flux cancelation triggers the minifilament jet eruptions) remained open. In \cite{panesar16b}, we found in on-disk quiet regions that coronal jets originate at a neutral line between dominant-polarity flux and a patch of canceling minority-polarity flux. Because we expect to have this same arrangement of canceling  dominant-polarity and minority-polarity flux at each jet base in coronal holes, we expect that on-disk coronal-hole jets work in the same fashion as on-disk quiet-region coronal jets. To test this hypothesis, here we investigate the on-disk coronal jet eruptions.

In this paper, we study the triggering mechanism of 13 randomly selected on-disk coronal-hole jets, using images from the \Hinode/X-ray telescope (XRT) and from the \textit{Solar Dynamics Observatory (SDO)}/Atmospheric Imaging Assembly (AIA), and using photospheric magnetograms from the \sdo/Helioseismic and Magnetic Imager (HMI). We systematically track the photospheric magnetic field evolution that leads to coronal-hole jets. We find that flux cancelation is the main process that triggers the coronal-hole jet eruptions, as in quiet-region jets.

\begin{figure*}
	\centering
	\includegraphics[width=\linewidth]{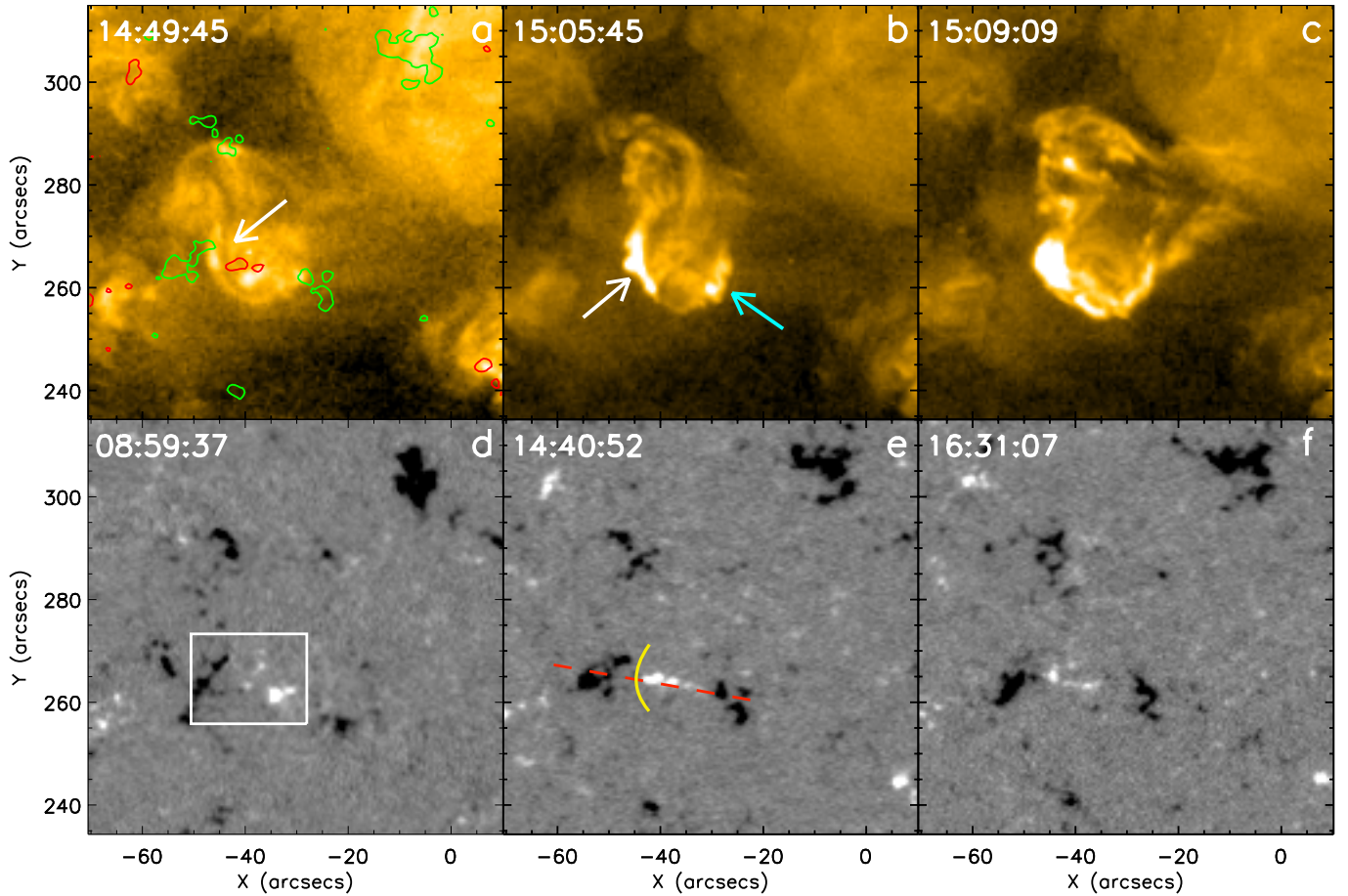}\vspace{-0.9cm}
	\caption{Coronal hole jet observed on 2017 Jan 03 (J11; Table \ref{tab:list}): Panels (a) - (c) show 171 \AA\  AIA intensity images. Panel (a) is prior to the eruption onset; the white arrow points to the pre-jet minifilament. Panels (b) and (c) show the eruption onset; in (b) the white and turquoise arrows point to the JBP and brightening from external reconnection, respectively. Panels (d) - (f) show line-of-sight HMI magnetograms of the jet region. The white box in (d) shows the region used to measure the positive magnetic flux for the plot shown in Figure \ref{fig3}a. The yellow line in (e) roughly outlines the strongest segment of the magnetic neutral line; it is where the JBP occurs during the eruption (see Figure \ref{fig2}). The red dashed line in (e) shows the location of the HMI time-distance map in Figure \ref{fig3}b; HMI contours of 14:51:22 UT and of level $\pm$ 50 G are overlaid onto panel (a), where red and green representing positive and negative polarities, respectively. Animations (MOVIE1a and MOVIE1b) of this Figure are available.} \label{fig1}
\end{figure*}

\begin{figure}
	\centering
	\includegraphics[width=\linewidth]{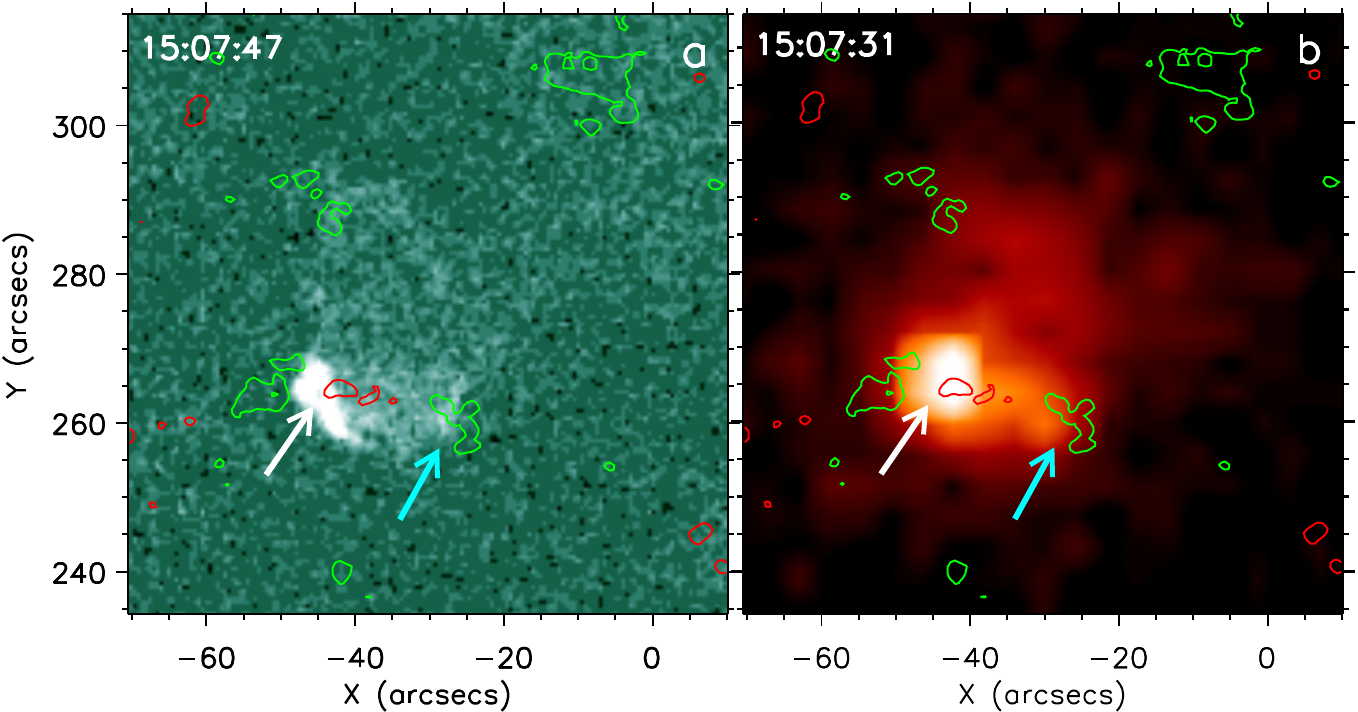}
	\caption{Jet Bright Point (JBP) of jet J11 of Table \ref{tab:list}: Panel (a) shows an 94 \AA\ AIA intensity image during the eruption onset. Panel (b) shows a concurrent  \Hinode/XRT image of the jet region. HMI contours of 15:07:07 UT and of level $\pm$ 50 G are overlaid onto panels (a) and (b), where red and green representing positive and negative polarities, respectively. The turquoise and white arrows in (a) and (b) point to the brightening from external reconnection and  to the (from internal reconnection) JBP that appears at the neutral line underneath the minifilament, respectively. } \label{fig2}
\end{figure} 
\section{INSTRUMENTATION AND DATA}\label{data} 
 \sdo/AIA provides full Sun images with high spatial resolution (0\arcsec.6 pixel$^{-1}$, $\sim$ 430 km) and temporal cadence (12 s) in seven EUV wavelength bands \citep{lem12}. For our investigations we use \sdo/AIA EUV images (304, 171, 193, 211 and 94 \AA) to view the transition-region-temperature and coronal-temperature jet plasma structures.
 
 To study the longer-term photospheric magnetic field evolution of the jet-base region, we use line-of-sight magnetograms from the \sdo/Helioseismic and Magnetic Imager (HMI; \citealt{schou12}), with  high spatial resolution of 0\arcsec.5 pixel$^{-1}$ and temporal cadence of 45 s \citep{scherrer12}.

 We also use \Hinode/XRT \citep{golub07} data, when available, data to view the coronal-temperature jet structures. Our XRT images have a limited field-of-view and a spatial resolution of 1\arcsec.02 pixel$^{-1}$, and with differing temporal cadences. XRT is sensitive to hot coronal emissions, detecting features of temperature $\gtrsim$ 1.0 $\times$ 10$^{6}$ K.
 
We randomly selected 13 on-disk coronal-hole jets by using AIA and XRT images from JHelioviewer software \citep{muller17}. All of the events and their measured parameters are listed in  Table \ref{tab:list}. Out of the 13 jets, only 6 of them  (Table \ref{tab:list}) were observed by XRT. Also, two jets (J9 and J10) were located at the boundary of a coronal-hole. The remaining 11 jets were fully inside of coronal-holes. In this paper, we will provide detailed analysis of two jets from our list of 13 jets. In Section \ref{results}, we show EUV and XRT observations for jet J11, and EUV observations for J1. The pre-jet magnetic field evolution of these two jets is discussed in Section \ref{flux}.

 \begin{figure*}
	\centering
	\includegraphics[width=\linewidth]{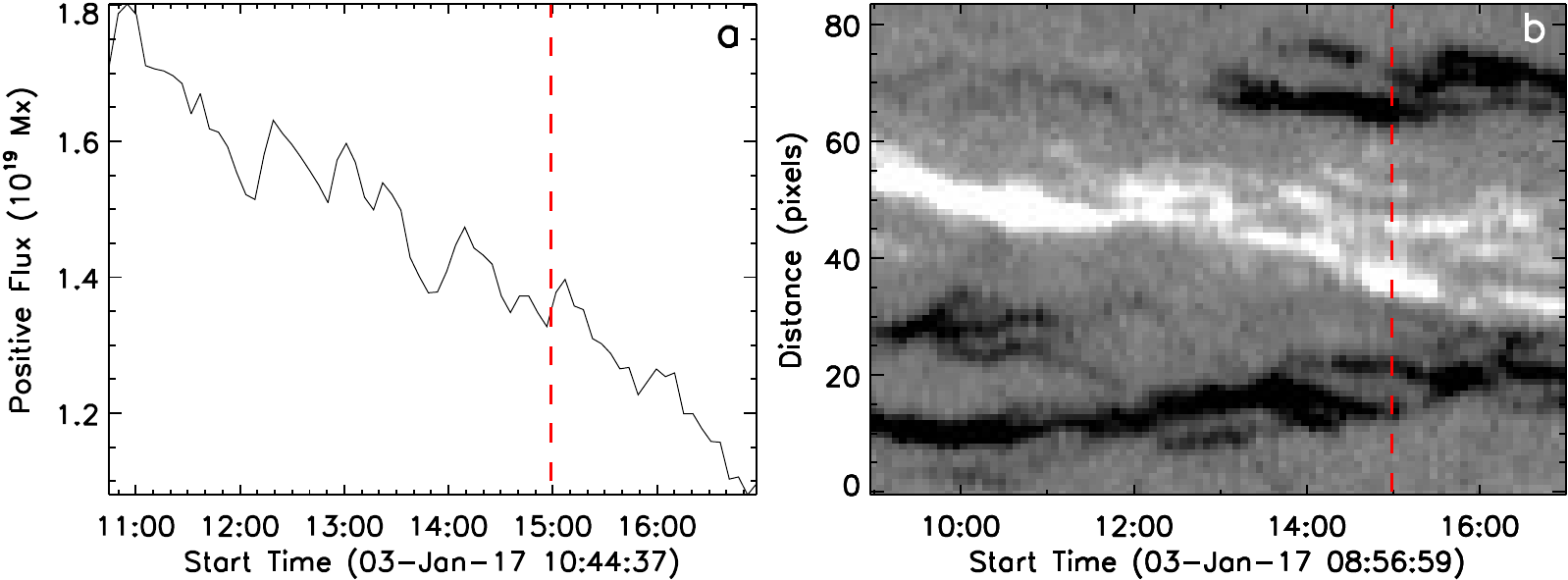}
	\caption{Magnetic flux cancelation J11 (Table \ref{tab:list}). Panel (a) shows the positive magnetic flux  as a function of time, integrated over the box region of Figure \ref{fig1}d. Panel (b) shows the HMI time-distance map along the red dashed line in Figure \ref{fig1}e. The red dashed lines in (a) and (b) show the jet-onset time.} \label{fig3}
\end{figure*} 

\section{Triggering of Minifilaments and Jets}\label{results}

\subsection{\textit{Jet (J11)}}\label{evo1}

Figures \ref{fig1}(a)-(c) show 171 \AA\ AIA images of the jet region with and without HMI contours. Figures \ref{fig1}(d)-(f) display the HMI photospheric magnetic field of the jet region. The accompanying movies (MOVIE1a and MOVIE1b) show the complete evolution of the pre-jet region. We first examine this pre-jet region in detail using AIA and XRT images. From the AIA images, we see that the minifilament was present at the neutral line at least $\sim$ 1-hr before the eruption onset. This coronal-hole has a dominant negative polarity, and prior to eruption (Figure \ref{fig1}) the minifilament resides on the neutral line (yellow line in \ref{fig1}e) between majority-polarity (negative) and minority-polarity (positive) flux clumps (Figures \ref{fig1}(a) and (d)).  From about 14:59 UT, the minifilament starts moving outward from the solar surface and then a JBP (white arrow in Figure \ref{fig1}b) appears underneath the rising minifilament. At the same time, a JBP appears in the XRT and 94 \AA\ AIA images (Figure \ref{fig2}).  After 15:04 UT, the minifilament is completely ejected and becomes a part of the jet spire (Figure \ref{fig1}c and MOVIE1a). The minifilament moves outward with an average speed of 105$\pm$30 \kms. During the rise of the minifilament external brightenings start to appear at the far-side of the majority-polarity (negative) flux clump. The AIA hotter channels (e.g. 94 \AA) and XRT images also show the signature of external brightenings at the majority-polarity negative flux clump (Figure \ref{fig1}(b) and Figure \ref{fig2}). This situation is broadly consistent with the polar coronal jet observations and schematic picture of \cite{sterling15}. The magnetic evolution of this jet is shown in Figure \ref{fig3}, and will be discussed in Section \ref{flux}.

\begin{figure*}
	\centering
	\includegraphics[width=\linewidth]{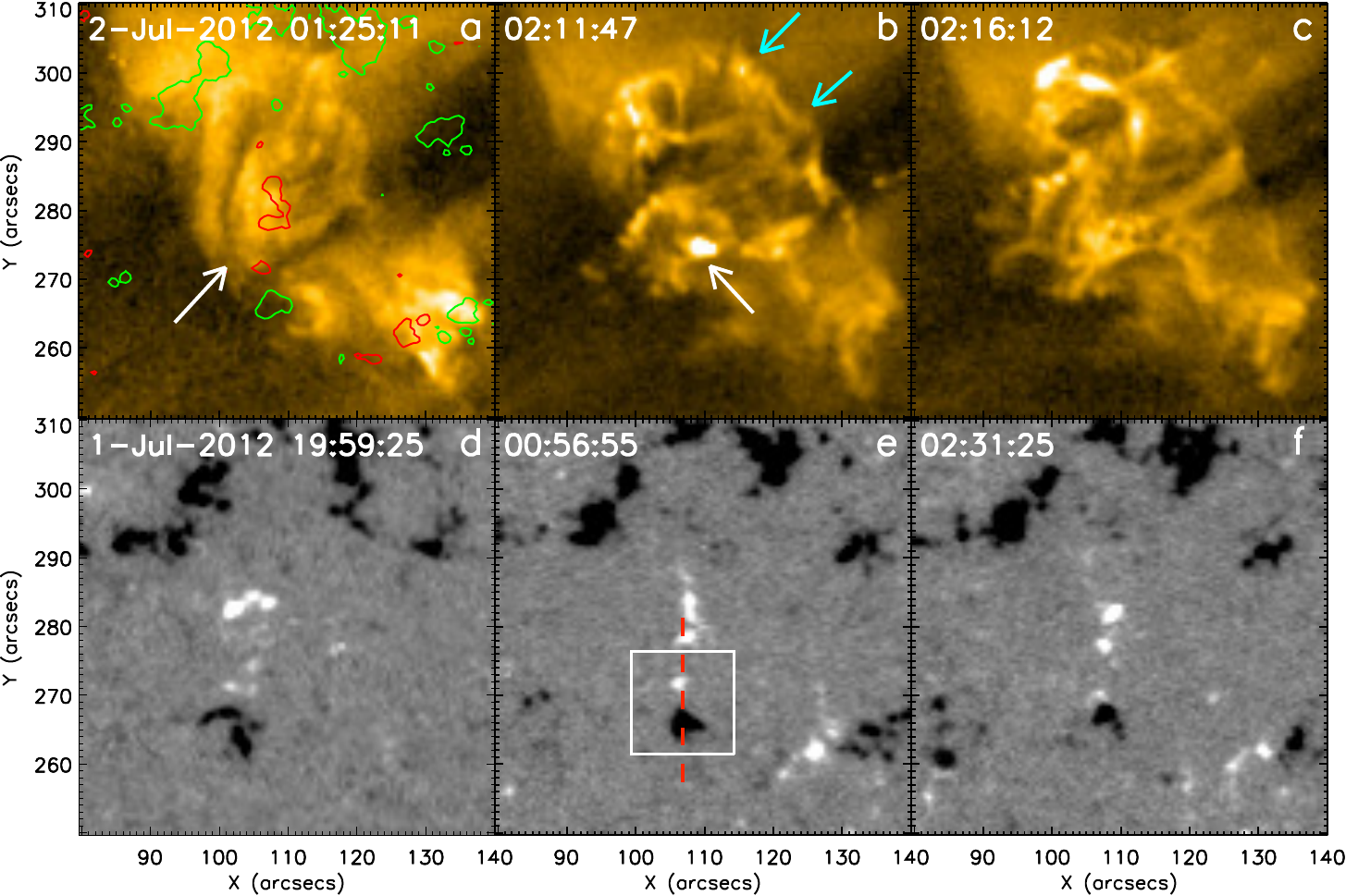} 
	\caption{Coronal hole jet observed on 2012 July 02 (J1 of Table \ref{tab:list}): The format is same as Figure \ref{fig1}. HMI contours of time 01:23:10 UT are overlaid onto the panel (a).  
		Animations (MOVIE2a and MOVIE2b) of this figure are available.}\label{fig4}
\end{figure*}

\subsection{\textit{Jet J1}}\label{evo2}

\begin{figure}
	\centering
	\includegraphics[width=\linewidth]{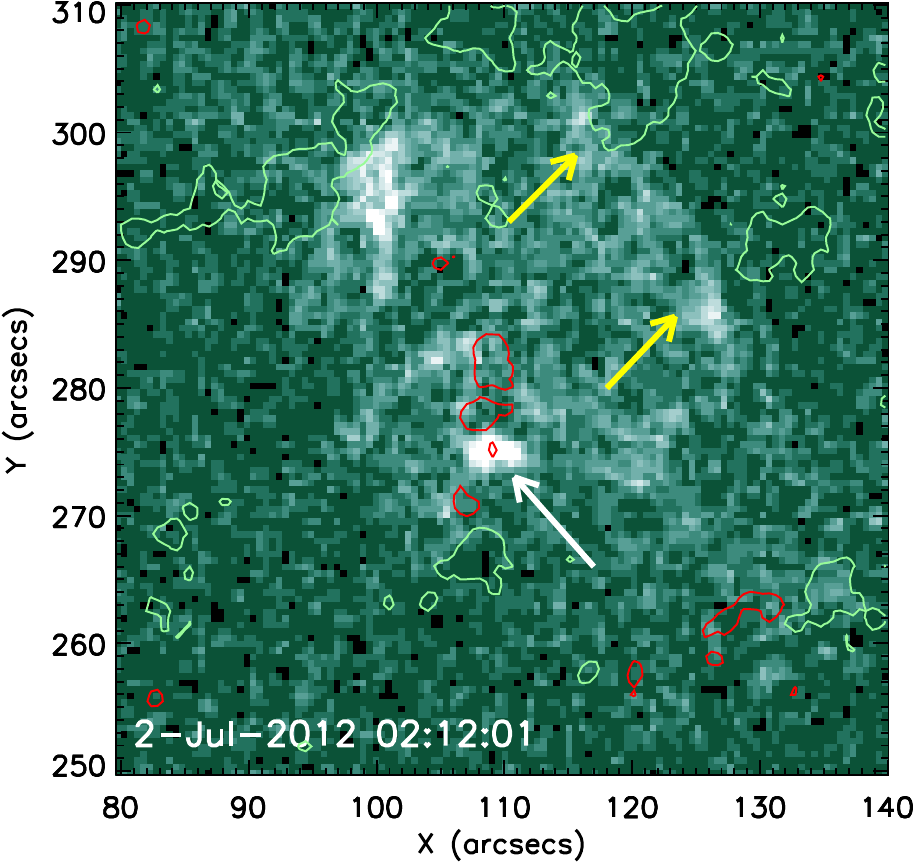}
	\caption{JBP of jet J1 of Table \ref{tab:list}, The format is same as Figure \ref{fig2}a.  HMI contours of time 02:10:25 and of level $\pm$ 50 G are overlaid onto the image. } \label{fig5}
\end{figure}

\begin{figure*}
	\centering
	\includegraphics[width=\linewidth]{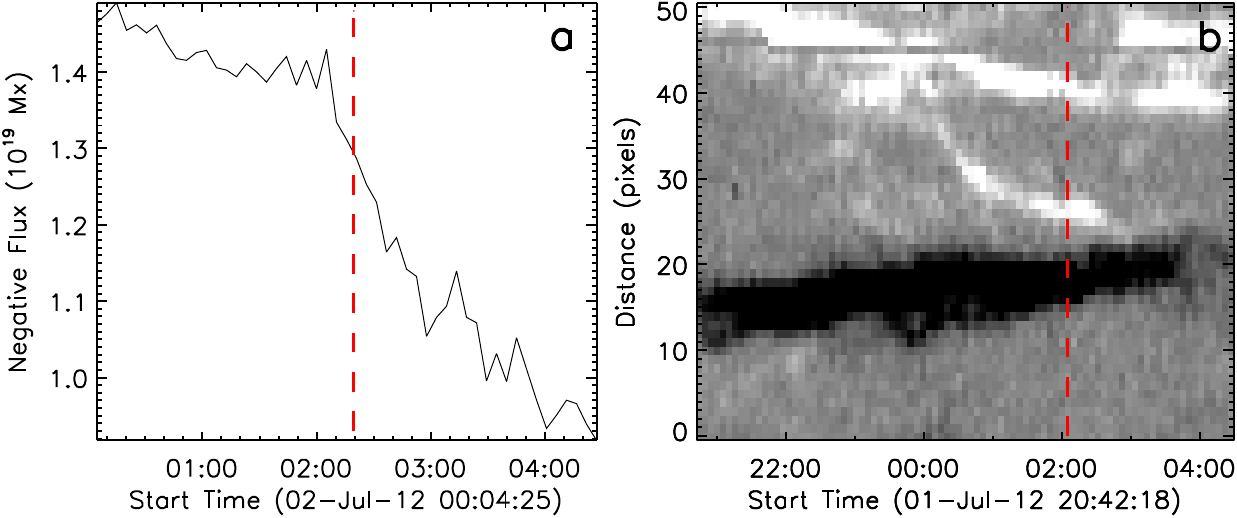}
	\caption{Magnetic flux cancelation J1 (Table \ref{tab:list}). The format is same as Figure \ref{fig3}.} \label{fig6} 
	
\end{figure*}

In Figure \ref{fig4}, we show our second detailed example of a jet, J1 from Table \ref{tab:list}. The high cadence AIA movie (MOVIE2a) accompanying Figure \ref{fig4}(a)-(c) shows the evolution of the minifilament eruption and jet. 
This is also a dominant negative-polarity coronal-hole region. For this jet we do not have XRT coverage so in Figure \ref{fig5} we only show an AIA 94 \AA\ image during the minifilament eruption. 
The white arrow in Figure \ref{fig4}(a) points to the minifilament, which clearly lies along a neutral line between majority-polarity (negative) and minority-polarity (positive) flux clumps (Figures \ref{fig4}(d)-(f)). The pre-jet minifilament was seen to be at this neutral line for more than $\sim$ 12 hrs. A noticeable rising of the minifilament body is underway at 02:05 UT. After that the minifilament erupts abruptly from the solar surface, with an average speed of 65$\pm$1.5 \kms, and the JBP turns on. Figure \ref{fig4}(b) and Figure \ref{fig5} show  that the JBP sits the neutral line where the minifilament was rooted before the eruption. Within 10 minutes, the AIA 171 \AA\ images show that the cool minifilament  material has been completely ejected along a broad spire. Thus this jet, and also the other remaining jets of Table \ref{tab:list},  consistent with the definition of blowout jets given by \cite{moore10,moore13}. This jet has the largest jet-base width (22500$\pm$1500 km, Table \ref{tab:list}) among the 13 jets. Moreover, a huge external brightening also appears at the (negative) majority-polarity flux clump (shown by the turquoise arrows in Figure \ref{fig4}b and yellow arrows in Figure \ref{fig5}). The magnetic evolution of this jet is shown in Figure \ref{fig6}, and  will be discussed in Section \ref{flux}.

\section{Magnetic Flux cancelation under the Minifilaments}\label{flux}

Figures \ref{fig1}(d)-(f) present line of sight magnetograms of the jet J11 region: (d) $\sim$ 6 hr before the eruption onset, (e) during the eruption onset, and (f) after the eruption onset of jet J11. 
The HMI magnetograms and movie (MOVIE1b) clearly show flux convergence and cancelation at the neutral line throughout the observation period. 

To see the longer-term evolution of the magnetic flux quantitatively we measure the (positive) minority-polarity flux of the jet region (J11) as a function of time, in the white box region of Figure \ref{fig1}d. Here, and with all our jets, we only measured the minority-polarity flux because it could be well isolated to avoid flows of flux of the selected polarity across the boundaries of the box. Figure \ref{fig3}a shows the integrated magnetic flux curve as a function of time. It shows a trend of continuous decrease in the positive flux for $\sim$ 6 hr, which is clear evidence of flux cancelation. The persistent flux cancelation at the neutral line finally triggers the minifilament eruption at 14:59 UT; the eruption time is indicated with the red dashed line in Figure \ref{fig3}a. We notice that in weak field regions, tiny grains of flux coalesce to make dense flux clumps; these likely are the cause of the  small bumps in the flux curve (e.g. at 12:10 UT).

Figure \ref{fig3}b displays the magnetic field evolution of the jet region over a time period of  $\sim$ 8 hr, as a time-distance map of the magnetic field along the red dashed line of Figure \ref{fig1}e. One can clearly see that the two polarities converge towards the neutral line and cancel with each other. In addition, there are some weak flux clumps close to the neutral line (that are not visible in the time-distance map) where flux cancelation goes on throughout the time period (e.g. at 10:49, 11:47, and 12:24 UT in MOVIE1b). Apparently, this continuous flux cancelation destabilizes the field holding the minifilament, resulting in its eruption. This behavior is in agreement with the quiet region jet eruptions presented by \cite{panesar16b}.

We can see in Figures \ref{fig3}a and \ref{fig3}b that the flux continues to cancel even after the jet eruption; it goes on until the minority-polarity flux has completely disappeared. In cases where minority-polarity flux remains after a jet and continues to cancel, there is a possibility of reformation of the minifilament. For example in jet J3 the eruption occurs at 09:54 UT, and after that flux continued to cancel, resulting in the reformation/reappearance of the minifilament at the same neutral line and a second eruption at 11:30 UT (the second jet eruption is not included in Table \ref{tab:list}). The jetting stops only when the minority-polarity flux patch has fully canceled \citep{panesar17}.

Figure \ref{fig6}a shows the integrated negative flux curve as a function of time for jet (J1). In this case we have measured the negative flux that was embedded inside the white box region of Figure \ref{fig4}e. There is a continuous drop in the negative flux, especially over 02:00 - 03:00 UT (Figure \ref{fig6}a and MOVIE2b), which is a clear indication of flux cancelation between the opposite polarity flux patches. Similarly, the flux cancelation is apparent in the HMI time-distance flux map of Figure \ref{fig6}b, leading to the minifilament eruption and jet onset at 02:11 UT (red dashed line of Figure \ref{fig6}).

\subsection{Flux Reduction}\label{red}

Each of the 13 coronal jets clearly shows  progressive  flux cancelation, and this apparently triggers the minifilament eruptions leading to the jets.  The opposite-polarity flux patches were seen to approach the neutral line beginning several hours (at least $\sim$ 3-4 hours, in some cases more than 3-4 hours e.g. in jet J11) before the eruptions. In order to obtain an estimate of the flux-reduction percentage for each eruption, we measured the average flux values 3-4 hours before the eruption and 0-2 hr after the eruption (Table \ref{tab:list}). We find that the flux clumps cancel with a flux loss of 20\% - 75\% from before to after the jet eruption. For quiet-region jets \citep{panesar16b}, the flux reduction was similar, 20\% - 60\% from before to after the jet eruption. 

We also calculated the average cancelation rate for our 13 events and found it to be  $\sim$ 0.6 $\times$ 10$^{18}$ Mx hr$^{-1}$. We can compare this with the quiet region jets of \cite{panesar16b} by revisiting the flux data of \cite{panesar16b}. We examined flux plots of all events of Table 1 of that paper, using the same procedure as above and we find that the average cancelation rate is $\sim$ 1.5 $\times$ 10$^{18}$ Mx hr$^{-1}$ for quiet region jets. In active region jet eruptions the flux cancelation rate was higher yet, $\sim$ 1.5 $\times$ 10$^{19}$ Mx hr$^{-1}$ \citep{sterling17}.

\section{SUMMARY AND DISCUSSION}\label{discussion}

We have examined the triggering mechanism of 13 on-disk coronal-hole jets. In each of the 13 events we find that a minifilament is present at a neutral line where the jet occurs, and that flux cancelation at that neutral line apparently triggers the minifilament eruption  driving the jet. These observations confirm that ondisk coronal-hole jets behave similar to ondisk quiet-region jets, which are also seen to erupt due to  continuous flux cancelation at a neutral line underneath a minifilament that erupts to drive them \citep{panesar16b}. Thus the schematic proposed by \cite{panesar16b} is also valid for the coronal-hole pre-jet minifilament eruptions because the overall idea is the same.

\begin{figure*}
	\centering
	\includegraphics[width=\linewidth]{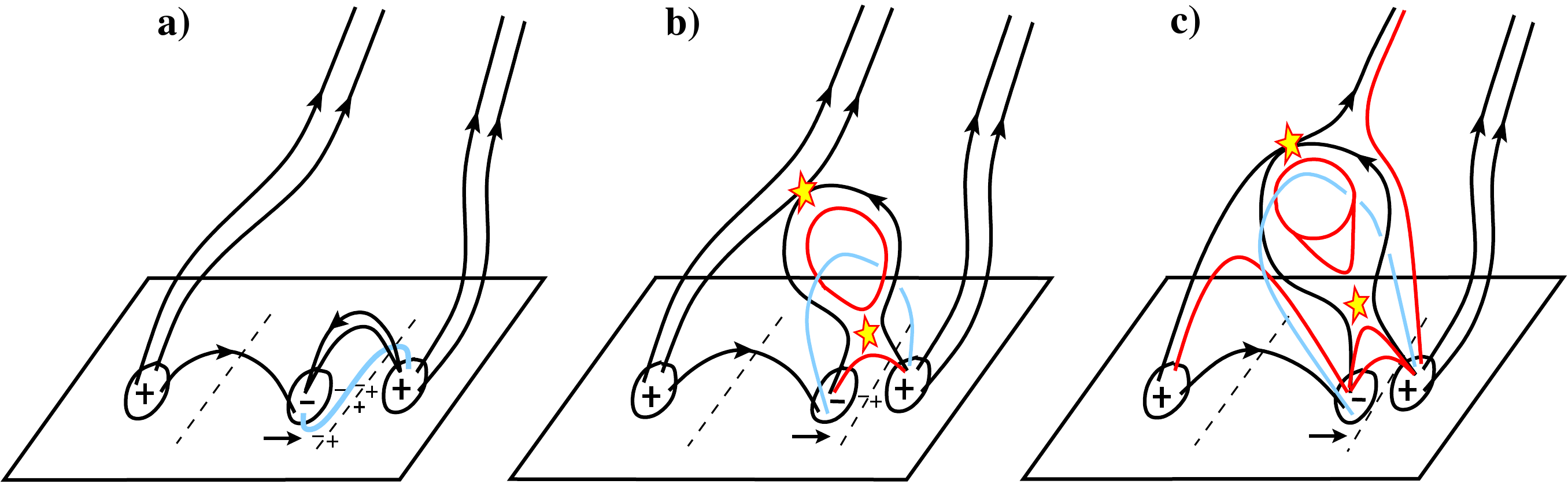}
	\caption{Schematic interpretation of the trigger of minifilament eruptions, as proposed in \cite{panesar16b}: this is a slightly modified picture of that in \cite{panesar16b}. The black rectangular box  represents the solar surface, the dashed lines show magnetic neural lines that separate the opposite-polarity flux patches, and the black and red lines respectively represent pre-existing magnetic field lines and newly-reconnected field lines. The positive and negative flux polarities are labeled with `+' sign and `-' sign, respectively. The blue  sigmoid-shaped structure in (a) represents a minifilament that sits on the right-hand-side neutral line, between a (negative) minority-polarity flux clump and a (positive) majority-polarity flux clump. The black loop above the minifilament is an overlying arcade. The black arrows near the bottom-right of the square in each panel indicates that minority-polarity flux is converging towards the neutral line and canceling with the opposite field. The thin `+' and `-' signs represent weaker canceling flux grains (as seen in the HMI movies). The lower and upper stars in (b) and (c) represent internal and external driven reconnection, respectively. The open red line shows the newly-reconnected open (or far-reaching closed) field line along which the jet material flows out.} \label{fig7}
\end{figure*}

Figure \ref{fig7} shows an embellished version of the schematic from \cite{panesar16b} for the trigger of quiet region pre-jet minifilament eruptions. A minority-polarity (negative field) flux patch resides in the dominant majority-polarity (positive) flux region and a minifilament (shown in blue color) sits at the neutral line of a small bipole (right hand side). The small (explosive) bipole contains a highly sheared and twisted field that holds the cool-minifilament material.
In Figure \ref{fig7}a, the minority-polarity and majority-polarity flux patches are well separated from each other, as in the HMI magnetograms well before the jet occurrence (Figures \ref{fig1} and \ref{fig4}). But  flux cancelation is already occurring continuously among weak flux grains that are closer to the neutral line (this is the embellishment of the schematic of \citealt{panesar16b}) than are the large flux patches. Eventually, the large flux patches also start to converge, and the continuous flux convergence and cancelation at the neutral line eventually destabilizes the field that carries the minifilament material and it erupts outward. This drives the \textit{internal reconnection} (lower star in (b) and (c)) that occurs in the legs of the erupting minifilament field, forming a JBP (low-lying red loop in (b) and (c)). The JBP appears at the location of the minifilament before the eruption (also see Figures \ref{fig2} and \ref{fig5}). The outer envelope of the erupting minifilament field undergoes driven reconnection (known as \textit{external/interchange} reconnection) with the impacted oppositely-directed open (or far-reaching) coronal field lines (upper star in (b) and (c)). The external reconnection produces two new magnetic connections: the red closed loop over the large bipole in (c) and the red open field line in (c). Hot material  (heated in the reconnection process)  and cool minifilament material escape along the newly-reconnected open field lines and appear as the jet spire. The EUV and X-ray images also show the signature of external brightenings at the far end of the majority-polarity flux clump (Figures \ref{fig2} and \ref{fig5}).

We now consider the origin of the minority-polarity flux patch that undergoes cancelation at the neutral line to build the magnetic field holding the minifilament. We find that either flux coalesces to form the minority flux clump; or there is an emergence of new bipole, and the minority-polarity foot of the bipole  starts canceling with nearby pre-existing majority-polarity flux. The continuous cancelation between the opposite-polarity flux patches and flux grains eventually destabilizes the field holding the minifilaments resulting in the minifilament eruption that generates the jet. This matches the picture presented by  \cite{moore92} for the triggering of  filament eruptions in flares and CMEs by flux cancelation at the filament's neutral line. Often, flux continues to cancel even after the jet occurs and only stops when the minority flux patch has completely canceled. These results are consistent with other recent  observations of jets resulting from magnetic flux cancelation \citep[e.g.][]{hong11,huang12,adams14,sterling16,sterling17}.

Among the six of our jets with XRT coverage, all except jet J9 have a corresponding EUV jet. Jet J9 is the only one that lacks a detectable corresponding EUV jet in any of the AIA channels; in the AIA images, only the JBP is visible. 
The observed jet speed (240$\pm$70 \kms, from XRT images) for J9 is higher than the speeds measured from AIA 171 \AA\ images for the remaining jets. That XRT-measured speed for jet J9 is in  agreement with the speeds of XRT coronal-hole jets observed by \cite{cirtain07}. We do not have enough temporal cadence for the remaining XRT jets to measure their speeds in the X-ray images.

The jets erupt outwards with an average speed and a standard deviation of 95$\pm$60 \kms\ (about the same as for the quiet region jets of \cite{panesar16b}) and the average duration of the studied jets is 8$\pm$3 minutes. The observed jet durations are similar to those of the coronal-hole jets of \cite{shimojo96} and \cite{savcheva07}. The average jet-base width is 11900$\pm$6000 km, which is marginally smaller than the quiet region jet-base widths (17000$\pm$5000\footnote{The uncertainty was misstated 
	in \cite{panesar16b} as $\pm$600 km; the correct value is 5000 km.} km)  obtained by \cite{panesar16b}.


Thus, our observations from this study show that flux cancelation is the trigger of coronal-hole jet eruptions. In summary, from our present study and previous studies \citep{panesar16b,panesar17}, we conclude that in coronal-holes, as in quiet regions, flux cancelation is the fundamental process for the buildup of the sheared and twisted magnetic field in and around the pre-eruption minifilament, and for the triggering the jet-driving minifilament eruption. 

\acknowledgments
N.K.P’s research was supported by an appointment to NASA Postdoctoral Program at the  NASA MSFC, administered by Universities Space Research Association under contract with NASA. A.C.S and R.L.M were supported by funding from the Heliophysics Division
of NASA's Science Mission Directorate through the heliophysics Guest Investigators Program, and by the \Hinode\ Project. We are indebted to the \sdo/AIA and \sdo/HMI teams for providing the high resolution data. \sdo\ data are courtesy of the NASA/\sdo\ AIA and HMI science teams.  \Hinode\ is a Japanese mission developed and launched by
ISAS/JAXA, with NAOJ as domestic partner and NASA and
STFC (UK) as international partners, and operated by these
agencies in co-operation with ESA and NSC (Norway).

\bibliographystyle{apj}

\end{document}